# Generalized Wilson loop method for nonlinear light-matter interaction

Hua Wang,[1,2,3] Xiuyu Tang,[1] Haowei Xu,[3] Ju Li,[3,4]* and Xiaofeng Qian[1,5]*

[1]Department of Materials Science and Engineering, Texas A&M University, College Station, TX 77843, USA

[2]Future Science Research Institute, ZJU-Hangzhou Global Scientific and Technological Innovation Center, Zhejiang University, Hangzhou, Zhejiang 310058, China

[3]Department of Nuclear Science and Engineering, Massachusetts Institute of Technology, Cambridge, Massachusetts 02139, USA

[4]Department of Materials Science and Engineering, Massachusetts Institute of Technology, Cambridge, Massachusetts 02139, USA

[5]Department of Electrical and Computer Engineering, Texas A&M University, College Station, TX 77843, USA

*Correspondence to: liju@mit.edu, feng@tamu.edu

**Abstract**

Nonlinear light-matter interaction, as the core of ultrafast optics, bulk photovoltaics, nonlinear optical sensing and imaging, and efficient generation of entangled photons, has been traditionally studied by first-principles theoretical methods with the sum-over-states approach. However, this indirect method often suffers from the divergence at band degeneracy and optical zeros as well as convergence issue and high computation cost when summing over the states. Here, using shift vector and shift current conductivity tensor as an example, we present a gauge-invariant generalized approach for efficient and direct calculations of nonlinear optical responses by representing interband Berry curvature, quantum metric, and shift vector in a generalized Wilson loop. This generalized Wilson loop method avoids the above cumbersome challenges and allows for easy implementation and efficient calculations. More importantly, the Wilson loop representation provides a succinct geometric interpretation of nonlinear optical processes and responses based on quantum geometric tensors and quantum geometric potentials, and can be readily applied to studying other excited-state responses.

## Introduction

Nonlinear light-matter interaction plays a pivotal role in ultrafast optics[1], bulk photovoltaics[2], nonlinear optical sensing and imaging[3], optical transistor[4], efficient generation of entangled photon pairs for quantum computing[5], etc. In particular, noncentrosymmetric materials are known to hold even-order nonlinear photocurrent responses under external electromagnetic field. For example, the wave packet of charge carriers can be displaced in real space upon photon excitation via a 2$^{nd}$ order process, resulting in shift current[2,6] that accounts for the shift mechanism for bulk photovoltaic effect.

Field-dependent nonlinear photocurrent can be obtained by solving the quantum kinetic equation of density matrix using perturbation theory. Subsequently, it can be calculated by first-principles methods such as density functional theory[7-10] and Wannier interpolation[11,12] with sum rules. However, the sum-over-states approach involves *ad hoc* cutoff that induces divergence at band degeneracy and optical zeros. Moreover, it suffers from the convergence issue with respect to the number of states. A direct approach is largely underexplored.

The Wilson loop method was originally proposed by Wilson[13] in 1974 for computing gauge field on a closed path. It is ubiquitous to gauge theories and has been widely used to calculate Berry curvature, Chern number, and other topological invariants in condensed matter physics, which are the hallmarks of a rich set of low-energy transport phenomena governed by the linear response of intraband process, including quantum Hall effect[14], quantum anomalous Hall effect[15], spin Hall effect[16], and quantum spin Hall effect[17,18]. Unlike the above linear responses, shift current involves interband transitions and its conductivity tensor is proportional to the quadratic electric field $E^2$. Young and Rappe[19] reformulated the shift vector using a gauge-invariant discrete expression similar to the King-Smith and Vanderbilt formalism of electric polarization[20]. Recently, Shi *et al.* represented the photon-drag shift vector with the Wilson loop formalism, important geometric quantity in shift current tensor and photon-drag shift current tensor[21]. These motivates us to develop a general approach for nonlinear optical (NLO) responses by representing interband Berry curvature, quantum metric, and shift vector in a generalized Wilson loop.

Here we present a physically intuitive gauge-invariant Wilson loop approach for direct and efficient calculations of NLO responses with Wilson loop representation, using the shift vector and shift current conductivity tensor as examples. In the Wilson loop picture, the geometrical nature of the shift current can be viewed as the difference of the spontaneous polarization determined by interband Berry connection between the valence and conduction bands upon direct optical transition. Unlike the standard sum-of-rule method, our Wilson loop approach is free of the convergence issue with respect to the number of states, and avoids the cumbersome divergence at band degeneracy and optical zeros where dipole matrix element is zero, that is, $r_{mn}^{a}(\boldsymbol{k}) \equiv \langle m|\hat{\boldsymbol{r}}_a|n\rangle = 0$. This quantum geometric approach can be easily implemented and allow for efficient calculations.

We demonstrate this powerful approach in two representative cases, including (1) a Rice-Mele model with an extra staggered onsite potential and (2) monolayer ferroelectric GeS with first-principles Wannier tight-binding Hamiltonian. The results calculated by the Wilson loop approach are in excellent agreement with the exact analytic solutions of the Rice-Mele model and the numerical results of monolayer GeS with the sum-over-states approach. Furthermore, we provide a Wilson representation of geometrical shift vector where the integral of the Wilson loop results in polarization difference between two bands upon optical transition, illustrating the geometrical nature of the shift current. In general, gauge-invariant geometric quantities, e.g., quantum metric, Berry curvature, and shift vector, can be all represented by Wilson loop naturally. The generalized Wilson loop approach developed here can be readily applied to other linear and nonlinear optical responses and allow for direct geometric interpretation of these quantities.

## Results

### Geometrical shift current response

Shift current originates from the difference of the real-space charge center of the valence and conduction bands upon optical transition. It is a bulk effect as the separation of photoexcited electrons and holes does not rely on *p*-*n* junction with built-in electric field. Unlike conventional photovoltaics, the generated open-circuit voltage can go beyond the bandgap, hence the power conversion efficiency is not limited by Shockley–Queisser limit for a single *p*-*n* junction. Under homogeneous linearly polarized light illumination, shift current can be written as[2,6,7,19]

$$J^a(0) = \sum_b 2\sigma^{abb}(0;\omega,-\omega)E^b(\omega)E^b(-\omega), \tag{1}$$

$$\sigma^{abb}(0;\omega,-\omega) = -\frac{\pi e^3}{\hbar^2}\int \frac{d\boldsymbol{k}}{(2\pi)^d}\sum_{nm} f_{nm}R_{mn}^{a,b}(\boldsymbol{k})r_{nm}^b r_{mn}^b\,\delta(\omega_{nm}-\omega), \tag{2}$$

$$R_{mn}^{a,b}(\boldsymbol{k}) = -\partial_{k_a}\phi_{mn}^b(\boldsymbol{k}) + \mathcal{A}_m^a(\boldsymbol{k}) - \mathcal{A}_n^a(\boldsymbol{k}), \tag{3}$$

where $\boldsymbol{r}_{nm} = i\langle n|\partial_{\boldsymbol{k}}|m\rangle$ for $n \neq m$ and $\boldsymbol{\mathcal{A}}_n = i\langle n|\partial_{\boldsymbol{k}}|n\rangle$ are interband and intraband Berry connection for states $|m\rangle$ and $|n\rangle$, respectively. $f$ is the Fermi-Dirac distribution with $f_{nm} \equiv f_n - f_m$. $\phi_{nm}^b(\boldsymbol{k})$ is the phase of Berry connection $r_{nm}^b(\boldsymbol{k})$ with $r_{nm}^b(\boldsymbol{k}) = |r_{nm}^b(\boldsymbol{k})|e^{i\phi_{nm}^b(\boldsymbol{k})}$. $R_{mn}^{a,b}(\boldsymbol{k})$ is the well-known shift vector described by the derivative of the phase and the difference of intraband Berry connection, also known as the quantum geometric potential[22]. Although the difference of Berry connections is gauge dependent, the shift vector is gauge invariant. We will discuss the gauge transformation property later.

The geometric aspect of the shift current is related with quantum metric and Berry curvature through the Christoffel symbols of the second kind[23,24]. The local quantum geometric tensor $Q_{mn}^{ab} = \langle\partial_a m|\tilde{Q}|\partial_b n\rangle$ was originally proposed by Provost and Vallee[25], where $\tilde{Q} = 1 - \tilde{P}$ and $\tilde{P}$ is the ground-state projection operator[26]. It indicates that the geodesic quantum distance between two quantum states in the Hilbert space can be viewed as absorption strength in the interband optical process, e.g., $Q_{mn}^{ab} = r_{mn}^a r_{nm}^b$. $Q_{mn}^{ab}$ consists of a symmetric quantum metric $g_{mn}^{ab}$ and an antisymmetric Berry curvature $\Omega_{mn}^{ab}$, *i.e.* $Q_{mn}^{ab} = g_{mn}^{ab} - \frac{i}{2}\Omega_{mn}^{ab}$. Quantum metric $g_{mn}^{ab}$ and Berry curvature $\Omega_{mn}^{ab}$ play quite different roles. For example, the off-diagonal $g_{mn}^{ab}$ and diagonal $\Omega_{mm}^{ab}$ contribute to the linear response coefficients of interband and intraband processes, respectively. In contrast, both quantum metric $g_{mn}^{ab}$ and Berry curvature $\Omega_{mn}^{ab}$ play a crucial role in second order responses. As we will show below, gauge-invariant geometric quantities such as quantum metric $g_{mn}^{ab}$, Berry curvature $\Omega_{mn}^{ab}$, and shift vector $R_{mn}^{a,b}$ can be all represented by Wilson loop naturally.

### Wilson loop approach of shift vector and shift current

Gauge-invariant single band Berry curvature in a discretized Brillouin zone can be calculated by Fukui-Hatsugai-Suzuki method[27], $\Omega_n^c(\boldsymbol{k}) = \mathrm{Im}\ln W_n(\boldsymbol{k})$, with

$$W_n(\boldsymbol{k}) = \epsilon_{abc}\langle n,\boldsymbol{k}|n,\boldsymbol{k}+\boldsymbol{q}_a\rangle\langle n,\boldsymbol{k}+\boldsymbol{q}_a|n,\boldsymbol{k}+\boldsymbol{q}_a+\boldsymbol{q}_b\rangle\langle n,\boldsymbol{k}+\boldsymbol{q}_a+\boldsymbol{q}_b|n,\boldsymbol{k}+\boldsymbol{q}_b\rangle\langle n,\boldsymbol{k}+\boldsymbol{q}_b|n,\boldsymbol{k}\rangle, \tag{4}$$

where $\boldsymbol{q}_a$ is an infinitesimal displacement vector along the corresponding *a* direction near $\boldsymbol{k}$ point and $\epsilon_{abc}$ is the Levi-Civita symbol. Now we derive a Wilson loop formula for the shift current response by reformulating shift vector with a strategy similar to that of King-Smith and Vanderbilt[20]. The intraband Berry connection reads $\boldsymbol{\mathcal{A}}_m(\boldsymbol{k}) = i\langle m,\boldsymbol{k}|\partial_{\boldsymbol{k}}|m,\boldsymbol{k}\rangle = \lim_{\boldsymbol{q}\to 0}\partial_{\boldsymbol{q}}\mathrm{Im}\ln\langle m,\boldsymbol{k}+\boldsymbol{q}|m,\boldsymbol{k}\rangle$. In contrast, the

interband Berry connection between states $|m\rangle$ and $|n\rangle$ is given by $\boldsymbol{r}_{mn}(\boldsymbol{k}) = i\langle n|\partial_{\boldsymbol{k}}|m\rangle = |\boldsymbol{r}_{mn}(\boldsymbol{k})|e^{i\phi_{mn}}$, where $\phi_{mn}$ is the phase of interband Berry connection: $\phi^b_{mn}(\boldsymbol{k}) = \mathrm{Im}\ln\left(r^b_{mn}(\boldsymbol{k})\right)$. For small $\boldsymbol{q}$, $\langle m,\boldsymbol{k}|m,\boldsymbol{k}+\boldsymbol{q}\rangle = e^{-i\boldsymbol{q}\cdot\mathcal{A}_m(\mathbf{k})+O(\boldsymbol{q}^2)}$, and $\langle m,\boldsymbol{k}+\boldsymbol{q}|m,\boldsymbol{k}\rangle = e^{i\boldsymbol{q}\cdot\mathcal{A}_m(\mathbf{k})+O(\boldsymbol{q}^2)}$. Thus, shift vector can be reformulated as

$$R^{a,b}_{mn}(\boldsymbol{k}) = -\partial_{k_a}\phi^b_{mn}(\boldsymbol{k}) + \mathcal{A}^a_m(\boldsymbol{k}) - \mathcal{A}^a_n(\boldsymbol{k}) = -\partial_{k_a}\mathrm{Im}\ln\left(r^b_{mn}(\boldsymbol{k})\right) + \mathcal{A}^a_m(\boldsymbol{k}) - \mathcal{A}^a_n(\boldsymbol{k})$$

$$= -\partial_{k_a}\mathrm{Im}\ \ln\left(r^b_{mn}(\boldsymbol{k})\right) - \lim_{q_a\to 0}\partial_{q_a}\mathrm{Im}\ \ln\langle m,\boldsymbol{k}|m,\boldsymbol{k}+\boldsymbol{q}_a\rangle\langle n,\boldsymbol{k}+\boldsymbol{q}_a|n,\boldsymbol{k}\rangle. \tag{5}$$

The first term on the right-hand side can be evaluated using finite difference around $\boldsymbol{k}$ along direction $a$,

$$\partial_{k_a}\mathrm{Im}\ \ln\left(r^b_{mn}(\boldsymbol{k})\right) = \lim_{q_a\to 0}\partial_{q_a}\mathrm{Im}\ \ln\langle m,\boldsymbol{k}+\boldsymbol{q}_a|r^b|n,\boldsymbol{k}+\boldsymbol{q}_a\rangle. \tag{6}$$

Thus,

$$R^{a,b}_{mn}(\mathbf{k}) = -\lim_{q_a\to 0}\partial_{q_a}\mathrm{Im}\ln\langle m,\boldsymbol{k}|m,\boldsymbol{k}+\boldsymbol{q}_a\rangle\langle m,\boldsymbol{k}+\boldsymbol{q}_a|r^b|n,\boldsymbol{k}+\boldsymbol{q}_a\rangle\langle n,\boldsymbol{k}+\boldsymbol{q}_a|n,\boldsymbol{k}\rangle. \tag{7}$$

Since $\langle n,\boldsymbol{k}|r^b|m,\boldsymbol{k}\rangle$ does not depend on $\boldsymbol{q}$, we can rewrite $R^{a,b}_{mn}(\mathbf{k})$ as

$$R^{a,b}_{mn}(\boldsymbol{k}) = -\lim_{q_a\to 0}\partial_{q_a}\mathrm{Im}\ln\langle m,\boldsymbol{k}|m,\boldsymbol{k}+\boldsymbol{q}_a\rangle\langle m,\boldsymbol{k}+\boldsymbol{q}_a|r^b|n,\boldsymbol{k}+\boldsymbol{q}_a\rangle\langle n,\boldsymbol{k}+\boldsymbol{q}_a|n,\boldsymbol{k}\rangle\langle n,\boldsymbol{k}|r^b|m,\boldsymbol{k}\rangle. \tag{8}$$

We then arrive at the Wilson loop representation of shift vector $R^{a,b}_{mn}(\boldsymbol{k})$ as

$$R^{a,b}_{mn}(\boldsymbol{k}) = -\lim_{q_a\to 0}\partial_{q_a}\mathrm{Im}\ln\left[W_{mn}(\boldsymbol{k},\boldsymbol{q}_a,r^b,r^b)\right] = -\lim_{q_a\to 0}\partial_{q_a}\arg\left[W_{mn}(\boldsymbol{k},\boldsymbol{q}_a,r^b,r^b)\right], \tag{9}$$

where $W_{mn}(\boldsymbol{k},\boldsymbol{q}_a,r^b,r^b)$ denotes general absorption on a Wilson loop,

$$W_{mn}(\boldsymbol{k},\boldsymbol{q}_a,r^b,r^b) \equiv \langle m,\boldsymbol{k}|m,\boldsymbol{k}+\boldsymbol{q}_a\rangle\langle m,\boldsymbol{k}+\boldsymbol{q}_a|r^b|n,\boldsymbol{k}+\boldsymbol{q}_a\rangle\langle n,\boldsymbol{k}+\boldsymbol{q}_a|n,\boldsymbol{k}\rangle\langle n,\boldsymbol{k}|r^b|m,\boldsymbol{k}\rangle, \tag{10}$$

$W_{mn}(\boldsymbol{k},\boldsymbol{q}_a = 0,r^b,r^b) = r^b_{mn}(\boldsymbol{k})r^b_{nm}(\boldsymbol{k})$ yields linear absorption strength at $\boldsymbol{k}$ through direct interband transition. The Wilson loop can be generalized to

$$W_{mn}(\boldsymbol{k},\boldsymbol{q},r^a,r^b) = \langle m,\boldsymbol{k}|m,\boldsymbol{k}+\boldsymbol{q}\rangle\langle m,\boldsymbol{k}+\boldsymbol{q}|r^a|n,\boldsymbol{k}+\boldsymbol{q}\rangle\langle n,\boldsymbol{k}+\boldsymbol{q}|n,\boldsymbol{k}\rangle\langle n,\boldsymbol{k}|r^b|m,\boldsymbol{k}\rangle. \tag{11}$$

Herein, the interband Berry curvature contributing to nonlinear injection current[10] can also be represented by the Wilson loop

$$\Omega^{ab}_{mn}(\boldsymbol{k}) = 2\mathrm{Im}W_{mn}(\boldsymbol{k},\boldsymbol{q} = 0,r^a,r^b). \tag{12}$$

In fact, $W_{mn}(\boldsymbol{k},\boldsymbol{q} = \boldsymbol{0},r^a,r^b)$ defined on the Wilson loop is identical to quantum geometric tensor, and its real and imaginary part gives the symmetric quantum metric $g^{ab}_{mn}$ and antisymmetric Berry curvature $\Omega^{ab}_{mn}$ at finite crystal momentum $\boldsymbol{k}$, respectively.

We can further rewrite the shift current conductivity tensor using the Wilson loop representation as

$$\sigma^{abb}(\omega) = -\frac{\pi e^3}{\hbar^2}\int\frac{d\boldsymbol{k}}{(2\pi)^d}\sum_{nm} f_{nm}R^{a,b}_{mn}(\boldsymbol{k})r^b_{nm}r^b_{mn}\,\delta(\omega_{nm}-\omega)$$

$$= \frac{\pi e^3}{\hbar^2}\int\frac{d\boldsymbol{k}}{(2\pi)^d}\sum_{nm} f_{nm}\lim_{q_a\to 0}\partial_{q_a}\mathrm{Im}\ln\left[W_{mn}(\boldsymbol{k},\boldsymbol{q}_a,r^b,r^b)\right]W_{mn}(\boldsymbol{k},0,r^b,r^b)\,\delta(\omega_{nm}-\omega)$$

$$= \frac{\pi e^3}{\hbar^2}\int\frac{d\boldsymbol{k}}{(2\pi)^d}\sum_{nm} f_{nm}\,\mathrm{Im}\lim_{q_a\to 0}\partial_{q_a}W_{mn}(\boldsymbol{k},\boldsymbol{q}_a,r^b,r^b)\,\delta(\omega_{nm}-\omega). \tag{13}$$

We then obtain

$$\sigma^{abb}(\omega) = \frac{\pi e^3}{\hbar^2} \int \frac{d\boldsymbol{k}}{(2\pi)^d} \sum_{nm} f_{nm} \lim_{q_a \to 0} \frac{1}{q_a} \operatorname{Im} W_{mn}\left(\boldsymbol{k}, \boldsymbol{q}_a, r^b, r^b\right) \delta(\omega_{nm} - \omega). \tag{14}$$

The above equation is a Wilson loop formula for shift current, which is only dependent on the imaginary part of the Wilson loop. This formula avoids the ambiguity in the definition of the argument around optical zeros where $r^a_{mn}(\boldsymbol{k}) \equiv \langle \psi_m | \hat{\boldsymbol{r}}_a | \psi_n \rangle = 0$ and at band degeneracies where $E_m = E_n$ or $\omega_{nm} = 0$. The Wilson loop representation of Berry curvature and geometrical shift vector are illustrated in Figs. 1a and 1b. In fact, the Wilson loop approach provides an equivalent expression as the Young-Rappe formula[19], which involves a more complicated Wilson loop, including indirect optical transition matrix elements as shown in Supplementary Figure 1.

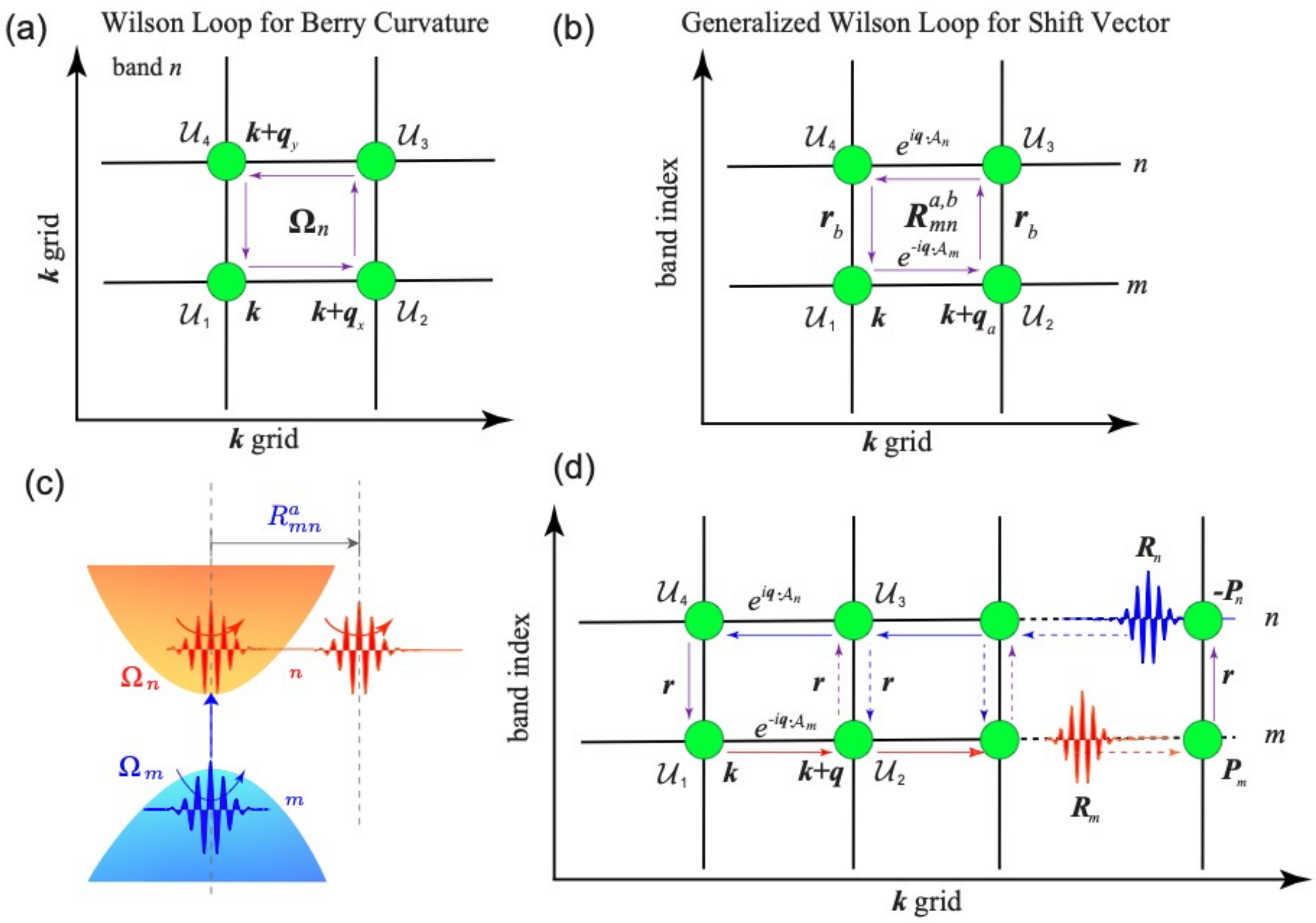

**Fig. 1 Wilson loop approach of the Berry curvature and geometrical shift vector. a** Berry curvature $\Omega^z(\boldsymbol{k})$ represented by the overlap matrix elements between Bloch wavefunctions $\mathcal{U}$ at neighboring $\boldsymbol{k}$-points. The Wilson loop is $W_n(\boldsymbol{k}) = \langle \mathcal{U}_1 | \mathcal{U}_2 \rangle \langle \mathcal{U}_2 | \mathcal{U}_3 \rangle \langle \mathcal{U}_3 | \mathcal{U}_4 \rangle \langle \mathcal{U}_4 | \mathcal{U}_1 \rangle$. **b** Shift vector $\boldsymbol{R}_{mn}(\boldsymbol{k})$ along $\boldsymbol{q}$ between band $m$ and $n$ represented by Bloch wavefunctions $\mathcal{U}$ at neighboring $\boldsymbol{k}$-points and transition matrix elements $\boldsymbol{r}_{mn}$. The Wilson loop for shift vector $W_{mn}(\boldsymbol{k}) = \langle \mathcal{U}_1 | \mathcal{U}_2 \rangle \langle \mathcal{U}_2 | \boldsymbol{r} | \mathcal{U}_3 \rangle \langle \mathcal{U}_3 | \mathcal{U}_4 \rangle \langle \mathcal{U}_4 | \boldsymbol{r} | \mathcal{U}_1 \rangle$ is associated with the interband transition. **c,d** Physical meaning of shift vector. The integral of the Wilson loops results in the polarization difference between two bands upon optical transition and the relative shift of the charge center of wave packet involving a pair of states.

Gauge invariance is guaranteed on each local Wilson loop in the discretized Brillouin zone. Under an arbitrary local gauge transformation $u_n(\boldsymbol{k}) \rightarrow e^{i\phi_n(\boldsymbol{k})} u_n(\boldsymbol{k})$ or $|n, \boldsymbol{k}\rangle \rightarrow e^{i\phi_n(\boldsymbol{k})} |n, \boldsymbol{k}\rangle$, Berry connections transform as

$$\boldsymbol{\mathcal{A}}'_m(\boldsymbol{k}) = \boldsymbol{\mathcal{A}}_m(\boldsymbol{k}) - \partial_{\boldsymbol{k}} \phi_m(\boldsymbol{k}), \tag{15}$$

$$\boldsymbol{r}'_{mn}(\boldsymbol{k}) = e^{i(\phi_n(\boldsymbol{k}) - \phi_m(\boldsymbol{k}))} \boldsymbol{r}_{mn}(\boldsymbol{k}). \tag{16}$$

The shift vector is clearly gauge-invariant

$$R'^{a,b}_{mn}(\boldsymbol{k}) = R^{a,b}_{mn}(\boldsymbol{k}) - \partial_{k_a}(\phi_n(\boldsymbol{k})) - \phi_m(\boldsymbol{k})) - \partial_{k_a}\phi_m(\boldsymbol{k}) + \partial_{k_a}\phi_n(\boldsymbol{k}) = R^{a,b}_{mn}(\boldsymbol{k}). \tag{17}$$

Consistently, the gauge transformation of quantum geometric tensor $W_{mn}(\boldsymbol{k},\boldsymbol{q},\boldsymbol{r},\boldsymbol{r})$ on the Wilson loop is given by

$$W'_{mn}(\boldsymbol{k},\boldsymbol{q},\boldsymbol{r},\boldsymbol{r}) = -\langle m,\boldsymbol{k}|m,\boldsymbol{k}+\boldsymbol{q}\rangle e^{i(\phi_m(\boldsymbol{k}+\boldsymbol{q})-\phi_m(\boldsymbol{k}))}\langle m,\boldsymbol{k}+\boldsymbol{q}|\boldsymbol{r}|n,\boldsymbol{k}+\boldsymbol{q}\rangle e^{i(\phi_n(\boldsymbol{k}+\boldsymbol{q})-\phi_m(\boldsymbol{k}+\boldsymbol{q}))}$$

$$\langle n,\boldsymbol{k}+\boldsymbol{q}|n,\boldsymbol{k}\rangle e^{i(\phi_n(\boldsymbol{k})-\phi_n(\boldsymbol{k}+\boldsymbol{q}))}\langle n,\boldsymbol{k}|\boldsymbol{r}|m,\boldsymbol{k}\rangle e^{i(\phi_m(\boldsymbol{k})-\phi_n(\boldsymbol{k}))}$$

$$= -\langle m,\boldsymbol{k}|m,\boldsymbol{k}+\boldsymbol{q}\rangle\langle m,\boldsymbol{k}+\boldsymbol{q}|\boldsymbol{r}|n,\boldsymbol{k}+\boldsymbol{q}\rangle\langle n,\boldsymbol{k}+\boldsymbol{q}|n,\boldsymbol{k}\rangle\langle n,\boldsymbol{k}|\boldsymbol{r}|m,\boldsymbol{k}\rangle = W_{mn}(\boldsymbol{k},\boldsymbol{q},\boldsymbol{r},\boldsymbol{r}). \tag{18}$$

Hence, quantum geometric tensor $W_{mn}(\boldsymbol{k},\boldsymbol{q},\boldsymbol{r},\boldsymbol{r})$ is also gauge invariant. Figure 1c shows the geometrical Berry curvature and shift vector using a two-band model. The geometrical meaning of the shift vector in Wilson loop representation is illustrated in Fig. 1d, which clearly shows it is related to the difference of the real-space charge center or spontaneous polarization for the valence and conduction bands upon direct optical transition. It is also known that the geometrical shift vector contributes to nonreciprocal Landau-Zener tunneling[28].

## Rice-Mele model of one-dimensional ferroelectric system

To demonstrate the generalized Wilson loop approach, we first use the one-dimensional Rice-Mele (RM) model of ferroelectric systems with broken inversion symmetry. The tight-binding Hamiltonian is illustrated in Fig. 2a, which is given by

$$H_{\mathrm{RM}} = \sum_i \left[\left(\frac{t}{2} + (-1)^i\frac{\delta_t}{2}\right)\left(c_i^\dagger c_{i+1} + h.c.\right) + (-1)^i\Delta_t c_i^\dagger c_{i+1}\right], \tag{19}$$

where $t$ is the hopping parameter and $\delta_t$ denotes the dimerization of the chain related to the distortion with respect to the centrosymmetric structure with $t_i = \frac{t}{2} + (-1)^i\frac{\delta_t}{2}$. $\Delta_t$ is the staggered on-site potential between two sites. $c_i^\dagger$ and $c_i$ are the fermion creation and annihilation operator, respectively. The inversion symmetry is broken when $\delta_t \neq 0$ and $\Delta_t \neq 0$. It leads to the following Bloch Hamiltonian

$$H_{\mathrm{RM}}(k) = \sum_{j=x,y,z} d_j\sigma_j = t\sigma_x\cos\left(\frac{ka}{2}\right) - \delta_t\sigma_y\sin\left(\frac{ka}{2}\right) + \Delta_t\sigma_z. \tag{20}$$

where $a$ is the lattice parameter. We use the following parameters for GeS, which yields a bandgap of 1.9 eV[9], $t = -1.0$eV, $\delta_t = -0.83$ eV, and $\Delta_t = -0.45$eV. The shift vector of the two-band RM model [29] has an analytical solution,

$$R_{cv} = \frac{\Delta_t a t\delta_t}{2E}\left\{\frac{(\delta_t^2 - t^2)[4E^2\cos ka + (t^2-\delta_t^2)\sin^2 ka]}{\Delta_t^2(\delta_t^2-t^2)\sin^2 ka + 4t^2\delta_t^2E^2} - \frac{1}{E^2-\Delta_t^2}\right\}, \tag{21}$$

The conduction and valence band energies $E_{v,c}$, as shown in Fig. 2b, are given by

$$E_{v,c} = \pm E = \pm\sqrt{t^2\cos^2\frac{ka}{2} + \delta_t^2\sin^2\frac{ka}{2} + \Delta_t^2}. \tag{22}$$

It is clear that the shift vector is reversed when $\delta_t$ or $\Delta_t$ changes sign, enabling ferroelectric-driven shift photocurrent switching[10]. This is also verified by Wilson loop approach as shown in Fig. 2d. Numerically, the shift vector and shift current conductivity are usually calculated by the sum rule with mass or diamagnetic term for tight-binding model. The generalized derivative of interband Berry connection for shift vector and shift current can be expressed by using the sum rule as[7,30]

$$r_{nm;b}^{a} = \frac{i}{\hbar\omega_{nm}}\left[ i r_{nm}^{a}\Delta_{nm}^{b} + i r_{nm}^{b}\Delta_{nm}^{a} - \hbar \sum_{p\neq n,m} \left( r_{np}^{a} r_{pm}^{b}\omega_{np} - r_{np}^{b} r_{pm}^{a}\omega_{pm} \right) - w_{nm}^{ab} \right], \quad (23)$$

where $\Delta_{nm}^{b} = v_{n}^{b} - v_{m}^{b}$ is the group velocity difference and $\hbar\omega_{nm} = \hbar\omega_{n} - \hbar\omega_{m}$ is the band energy difference. The mass term $w_{nm}^{ab} = \langle n | \partial_{k_a}\partial_{k_b} H | m \rangle$ cannot be neglected for the tight-binding model because the interband Berry connection is not gauge-covariant and its generalized derivative in the Hamiltonian gauge involves the second-order derivative of Hamiltonian [31]. In two-band RM model with $m = 1, n = 2$, the shift vector using the sum rule at optical nonzero $\boldsymbol{k}$-points (*i.e.* $|r_{nm}^{b} r_{mn}^{b}| \neq 0$) reads

$$R_{nm}^{a,b} \equiv \frac{\mathrm{Im}\, r_{nm;a}^{b} r_{mn}^{b}}{|r_{nm}^{b} r_{mn}^{b}|} = \frac{\mathrm{Re}\, w_{nm}^{ba} r_{mn}^{b}}{|r_{nm}^{b} r_{mn}^{b}| \hbar\omega_{mn}}. \quad (24)$$

It shows that the shift vector and shift current are vanishing in two-band RM model without considering the mass term. The calculated shift vector and shift current conductivity with an effective area of 9.37 Å$^2$ by different methods are shown in Figs. 2c and 2d. The results demonstrate that the shift vector and the shift current calculated by the Wilson loop approach are in excellent agreement with the analytical solution and the sum-over-states approach.

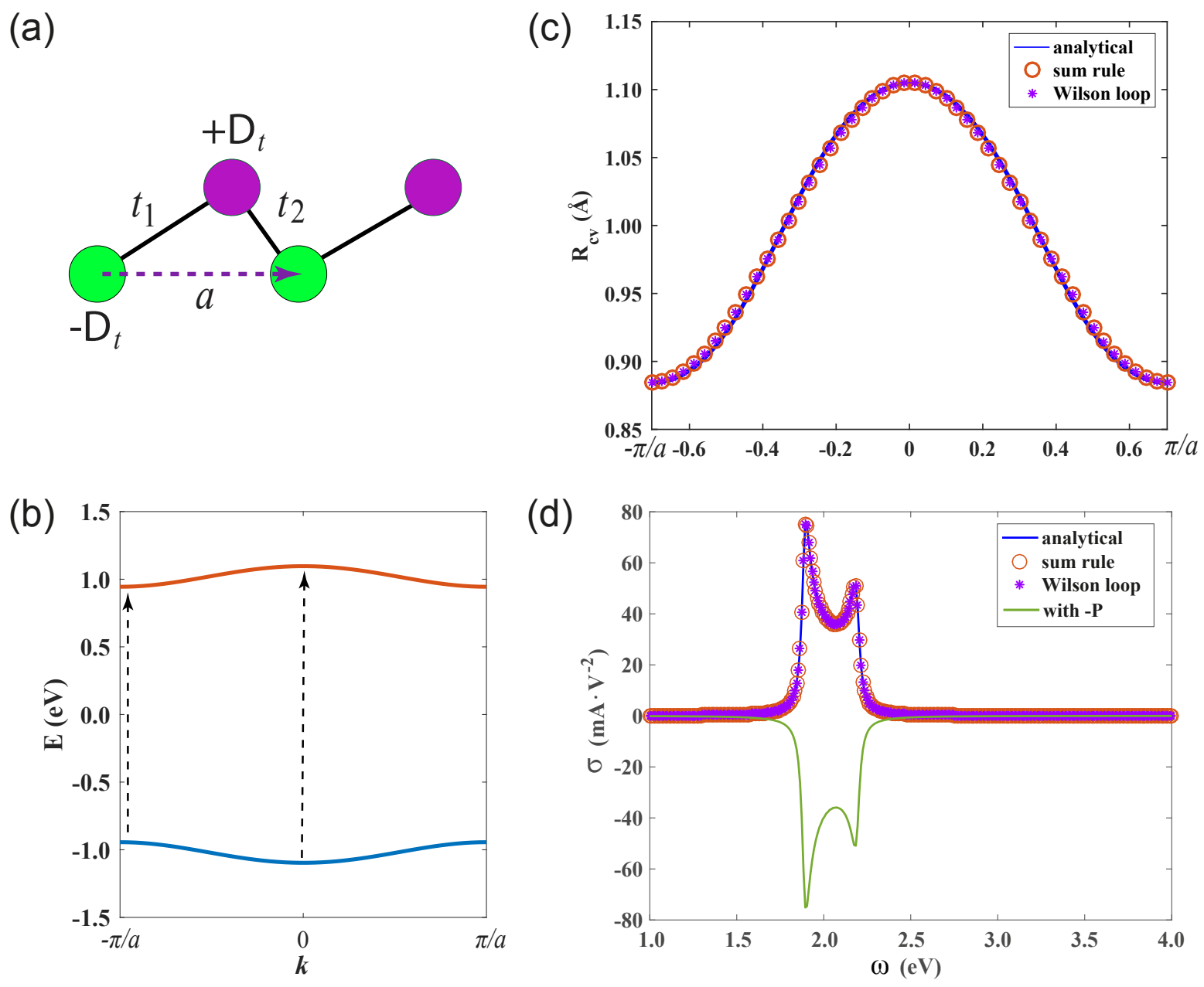

**Fig. 2 Band structure, shift current, and shift vector for noncentrosymmetric Rice-Mele model. a** Two-band Rice-Mele tight-binding model of one-dimensional polarized chain with two sites in each unit cell. **b** Energy dispersion of the conduction and valence bands. The arrows denote the optical transition in the edge and center of the first Brillouin zone. **c** Shift vector calculated by three different methods, including the analytical solution, the sum rule with the mass term, and the Wilson loop approach. **d** Shift current calculated by the analytical method, the sum rule with mass term, and the Wilson loop approach. "-P" denotes the 1D Rice-Mele model with reversed polarization. The two peaks are related to the optical transitions indicated as two arrows in **b**.

## Wannier tight-binding model of monolayer GeS

Next, we demonstrate the Wilson loop method for real materials with a symmetrized Wannier tight-binding Hamiltonian. The details of the first-principles calculations and the constructions of symmetrized Wannier tight-binding Hamiltonian are described in Methods. Taking the ferroelectric monolayer GeS as an example, it has a $C_{2v}$ point group with a mirror plane $\mathcal{M}_x$ perpendicular to the $x$-axis. The crystal structure and band structure of monolayer GeS are shown in Figs. 3a and 3b, respectively. From group theory analysis, the components $\sigma^{xxx}(\omega)$ and $\sigma^{xyy}(\omega)$ vanish under linearly polarized light with in-plane polarization, which was verified in our calculation. Here, we focus on $\sigma^{yyy}(\omega)$ and the corresponding shift vector $R_{cv}^{y,y}$. Figure 3c shows $\boldsymbol{k}$-resolved shift vector $R_{cv}^{y,y}$ between the top valence band and the bottom conduction band across the bandgap. The shift vector away from optical zeros can be ~10 Å, much larger than its lattice constant. This is very different from spontaneous electric polarization vector constrained within the lattice constant. Berry curvature of the top valence band is also calculated by the Wilson loop approach as shown in Fig. 3d. Given the mirror symmetry $\mathcal{M}_x$, we have verified the symmetry properties $R_{cv}^{y,y}(k_x, k_y) = R_{cv}^{y,y}(-k_x, k_y)$ and $\Omega_v^z(k_x, k_y) = -\Omega_v^z(-k_x, k_y)$. A Berry curvature dipole along $x$ direction is generated which can induce a similar ferroelectric nonlinear Hall effect[32-34] in monolayer GeS. The intraband Berry curvature of the bottom conduction band and the interband Berry curvature across the gap are presented in Supplementary Figure 2.

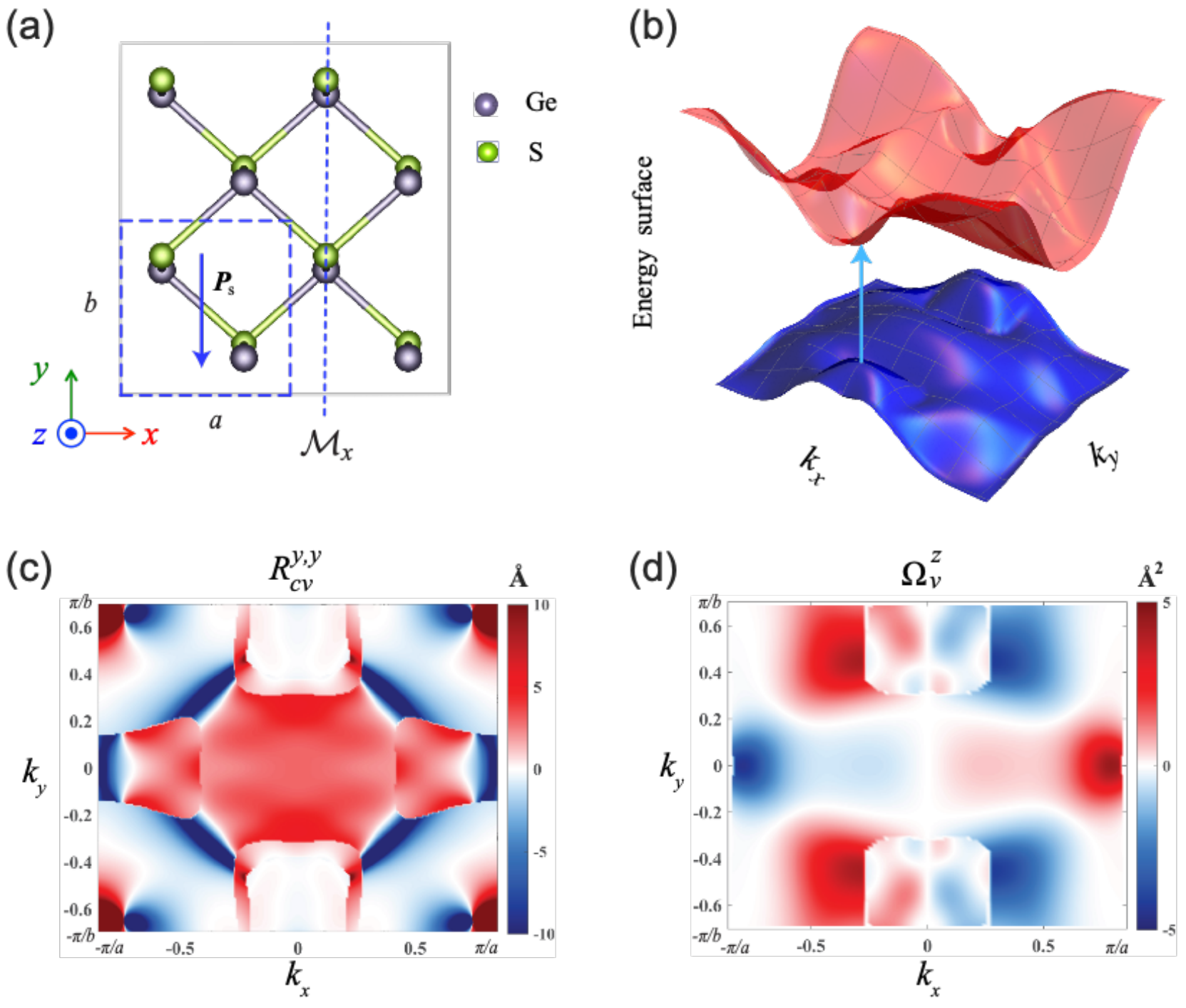

**Fig. 3 Crystal structure, band structure, shift vector, and Berry curvature of monolayer GeS. a** Crystal structure of ferroelectric monolayer GeS with noncentrosymmetric $C_{2v}$ point group. **b** Valence and conduction band energy surfaces across the bandgap. **c** $k$-resolved shift vector across the bandgap in the first Brillouin zone. **d** $k$-resolved Berry curvature of the top valence band in the first Brillouin zone.

Figure 4a shows the calculated frequency-dependent shift current conductivity $\sigma^{yyy}(\omega)$. To convert the sheet conductivity $\sigma^{2D}$ to bulk conductivity $\sigma^{3D}$, we set the effective thickness $l$ to be 2.56 Å by $\sigma^{3D} = \sigma^{2D}/l$. We have verified the identity $W_{mn}(\boldsymbol{k}, \boldsymbol{q} = 0, r^b, r^b) = r_{mn}^b r_{nm}^b$. The $k$-resolved absorption strength between the top valence band and the bottom conduction band is shown in Fig. 4b. The white region indicates optical zeros and has no contribution to shift current conductivity. To investigate the origin of large responses, we calculate the $k$-resolved shift current strength $I_{mn}^{a,b}(\boldsymbol{k},\omega)$ at $\omega = 2.0$ and $\omega = 2.8$ eV using the Wilson loop approach, defined as

$$I_{mn}^{a,b}(\boldsymbol{k},\omega) = \lim_{q_a \to 0} \frac{1}{q_a} \operatorname{Im} W_{mn}\big(\boldsymbol{k}, \boldsymbol{q}_a, r^b, r^b\big)\delta(\omega_{nm} - \omega). \tag{25}$$

The results are shown in Figs. 4c and 4d. The convergence was carefully checked with respect to the number of bands and $\boldsymbol{k}$-point sampling in the first Brillouin zone, as shown in Supplementary Figure 3, for shift current conductivity. The calculated shift current conductivity is well converged with a $\boldsymbol{k}$-point mesh of 200×200×1 and two bands below 3 eV. In addition, frequency-dependent $\sigma^{yxx}(\omega)$ for monolayer GeS is shown in Supplementary Figure 4. Furthermore, we performed similar calculations with the Wilson loop approach for a different 2D materials monolayer $WS_2$, and the results are shown in Supplementary Figure 5. Our results clearly demonstrate that the generalized Wilson loop approach is not only efficient and generally applicable to both effective models and realistic materials, but also avoids the summation over a large number of intermediate valence and conduction bands, making it valuable for computing nonlinear optical responses.

It should be noted that the geometrical shift vector at optical zeros cannot be calculated using the sum-over-states approach. Furthermore, while the large shift vector at optical zeros has no contribution to shift current response with vertical transitions, it can contribute to the shift current response when taking into account the photon-drag effect[21] involving indirect transitions or strong electron-phonon coupling effect. Our demonstration of geometrical shift vector in real materials will allow for theoretical investigations of a wide range of geometric effects induced by quantum geometric potential.

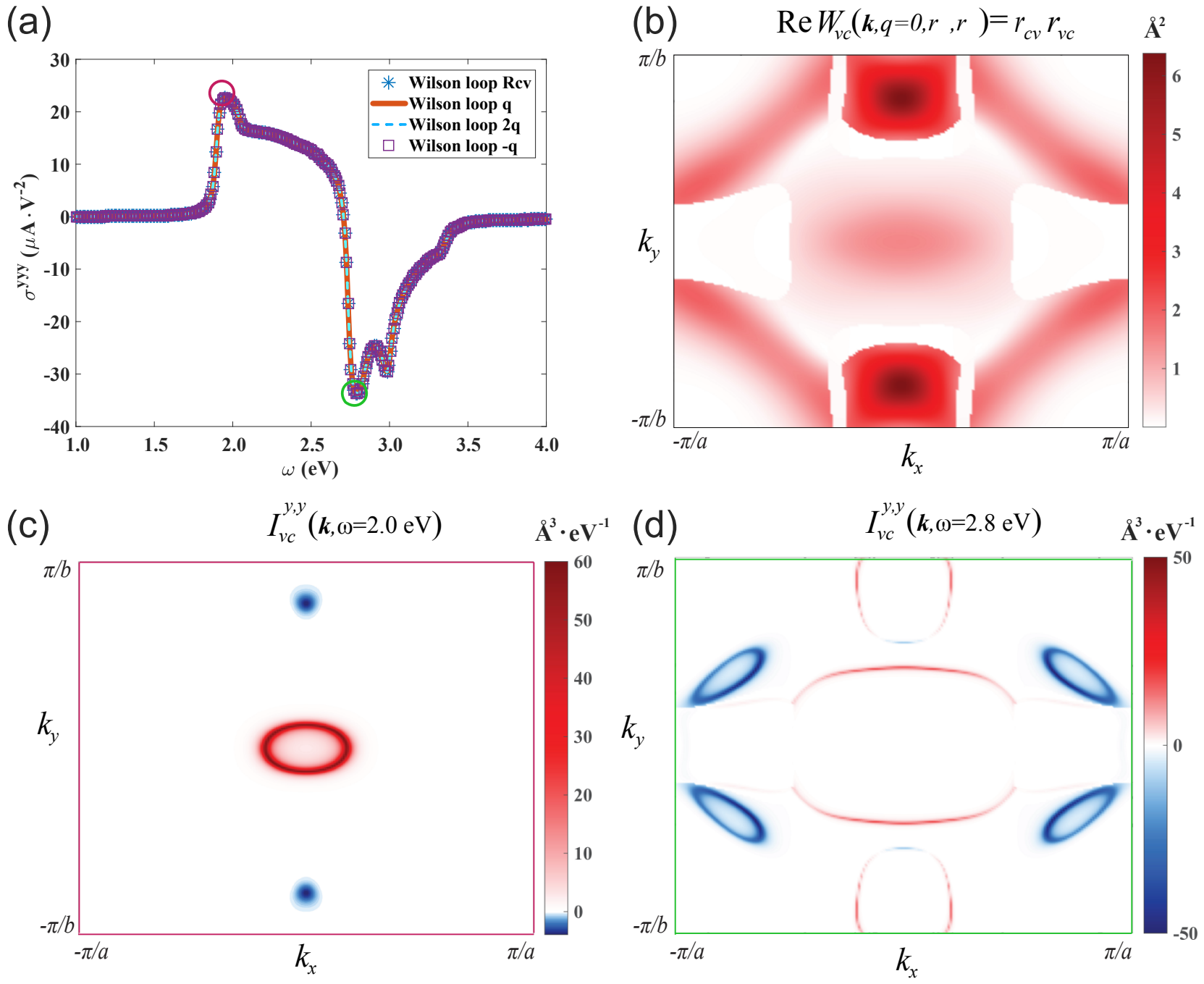

**Fig. 4 Shift current conductivity and the *k*-space distribution of the related quantities in monolayer GeS.** **a** Shift current conductivity $\sigma^{yyy}(\omega)$ calculated by the Wilson loop approach with two bands across the gap. $R_{cv}$ denotes the original shift vector formula and $q$, $2q$, $-q$ represent the formula using the imaginary part of Wilson loop $W_{mn}(\boldsymbol{k},\boldsymbol{q},\boldsymbol{r},\boldsymbol{r})$ with different $q$ values. **b** $k$-resolved absorption strength $r_{cv}^{y}r_{vc}^{y}$ corresponding to quantum metric $g_{cv}^{yy}$ in the first Brillouin zone. **c,d** $k$-resolved shift current strength $I_{vc}^{y,y}(\boldsymbol{k},\omega)$ at $\omega = 2.0$ and $\omega = 2.8$ eV using the Wilson loop approach, respectively.

## Disscussion

A common challenge with a perturbation theory for NLO responses in the length gauge is the treatment of the position operator $\boldsymbol{r}$ for the extended Bloch states. The intraband part of the position operator is represented by $\langle m, \boldsymbol{k}|\boldsymbol{r}_i|n, \boldsymbol{k}'\rangle = \delta_{mn}\big(\delta(\boldsymbol{k}-\boldsymbol{k}')\boldsymbol{\mathcal{A}}(\boldsymbol{k}) + i\partial_{\boldsymbol{k}}\delta(\boldsymbol{k}-\boldsymbol{k}')\big)$. The real-space coordinate $\boldsymbol{r}$ of the wave packet made from the Bloch wave functions is represented by $\boldsymbol{r}_i = i\partial_{\boldsymbol{k}} - \boldsymbol{\mathcal{A}}(\boldsymbol{k})$. NLO responses involve the matrix element of the commutator $\langle m, \boldsymbol{k}|[\boldsymbol{r}_i, \mathcal{O}]|n, \boldsymbol{k}'\rangle = i\delta(\boldsymbol{k}-\boldsymbol{k}')\mathcal{O}_{mn;\boldsymbol{k}}$, where the covariance derivative $\mathcal{O}_{mn;\boldsymbol{k}} \equiv \partial_k \mathcal{O}_{mn} - i\mathcal{O}_{mn}\big(\boldsymbol{\mathcal{A}}_m(\boldsymbol{k}) - \boldsymbol{\mathcal{A}}_n(\boldsymbol{k})\big)$ plays a central role in many other nonlinear responses. The derivation of the commutator relation can be found in Supplementary Information. For example, the generalized derivative of dipole matrix element is written as

$$(r_{nm}^a)_{;k^b} \equiv \partial_b(r_{nm}^a) - i\big(\xi_{nn}^b - \xi_{mm}^b\big)r_{nm}^a = -i\big[-\partial_b(\phi_{nm}^a) + \big(\mathcal{A}_n^b - \mathcal{A}_m^b\big)\big]r_{nm}^a = iR_{mn}^{a,b}r_{nm}^a, \qquad (26)$$

which is a key physical quantity for second harmonic generation[7,35]. Hence, the Wilson loop approach developed here can be readily applied to other linear and nonlinear optical effects such as second and third harmonic generation and linear and quadratic electro-optic effect.

The spin-orbit interaction is weak in monolayer GeS, thus it is not considered in the present calculations. Nevertheless, the expression can be easily extended to include spin-orbit coupling, and generalized to the degenerate and near degenerate cases by considering connected and disconnected band sets and using $\boldsymbol{k}\cdot\boldsymbol{p}$ perturbation theory to obtain a smooth variation of the Bloch states between $\boldsymbol{k}$ and $\boldsymbol{k}+\delta\boldsymbol{k}$[36]. Furthermore, although all the calculations in this work are performed within the independent particle approximation, the Wilson loop approach can also be developed to include many-body effect.

In summary, we presented a gauge-invariant generalized approach for efficient and direct calculations of nonlinear optical responses with pure Wilson loop representation. This generalized Wilson loop method avoids the cumbersome issues of the commonly used sum-over-states approach, and allows for easy implementation and efficient calculation. The Wilson loop representation provides an elegant geometric interpretation of nonlinear optical processes and responses based on quantum geometric tensors and quantum geometric potentials responsible for shift current and Landau-Zener tunneling. The generalized Wilson loop method developed here can be readily applied to study other nonlinear optical responses such as second and third harmonic generation, linear and quadratic electro-optic effect, as well as magnetic injection current and magnetic shift current[37].

## Methods

**First-principles calculations of atomistic and electronic structure.** First-principles calculations for structural relaxation and quasiatomic Wannier functions were performed using density-functional theory[38,39] as implemented in the Vienna Ab initio Simulation Package (VASP)[40] with the projector-augmented wave method for treating core electrons[41]. We employed the generalized-gradient approximation of exchange-correlation functional in the Perdew-Burke-Ernzerhof form[42], a plane-wave basis with an energy cutoff of 400 eV, and a Monkhorst-Pack $k$-point sampling of 10×10×1 for the Brillouin zone integration.

**Generalized Wilson loop approach of shift current using first-principles tight-binding Hamiltonian.** To compute the Wilson loop related quantities, we first construct quasiatomic Wannier functions and symmetrized first-principles tight-binding Hamiltonian from Kohn-Sham wavefunctions and eigenvalues under the maximal similarity measure with respect to pseudoatomic orbitals[43,44]. Total 16 quasiatomic Wannier functions were obtained for monolayer GeS. Using the developed tight-binding Hamiltonian we then compute Berry curvature, shift vector and shift current using a dense $k$-point sampling of 200×200×1. Sokhotski-Plemelj theorem is employed for the Dirac delta function integration with a small imaginary smearing factor $\eta$ of 0.02 eV. We checked the convergence of shift current conductivity tensor with respect to the number of bands and the $k$-point sampling as well as different q values of the reciprocal lattice used in the Wilson loop method. (see Supplementary Figure 3 and 4).

**Symmetrization of the tight-binding Hamiltonian.** The construction of Wannier functions for crystals does not preserve space group symmetries. To avoid the artificial symmetry breaking, we performed symmetrization of the tight-binding Hamiltonian. The Hamiltonian is invariant under symmetry operation $g$ in the group $G$

$$\forall g \in G: H(\boldsymbol{k}) = D_{\boldsymbol{k}}(g)H(g^{-1}\boldsymbol{k})D_{\boldsymbol{k}}(g^{-1}), \tag{27}$$

$$D_{\boldsymbol{k}}(g) = e^{i\boldsymbol{t}_g\cdot\boldsymbol{k}}D(g), \tag{28}$$

where $D_{\boldsymbol{k}}(g)$ is $\boldsymbol{k}$-dependent representation of the symmetry and $\boldsymbol{t}_g$ is the translation vector of the symmetry. We define the symmetrized Hamiltonian using the group average

$$\widetilde{H}(\boldsymbol{k}) = \frac{1}{|G|}\sum_{g\in G} D_{\boldsymbol{k}}(g)H(g^{-1}\boldsymbol{k})D_{\boldsymbol{k}}(g^{-1}). \tag{29}$$

To apply the group average to above tight-binding Hamiltonian with all crystalline symmetry constraints in real space, we rewrite the hopping matrices[45]

$$\widetilde{H}_{ij}(\boldsymbol{R}) = \frac{1}{|G|}\sum_{g\in G} D_{il}(g)H_{lm}\left(S_g^{-1}\left(\boldsymbol{R} - \boldsymbol{T}_{ij}^{ml}\right)\right)D_{mj}(g^{-1}), \tag{30}$$

where $S_g$ is the real space rotation matrix and $\boldsymbol{T}_{ij}^{ml} = S_g(\boldsymbol{r}_m - \boldsymbol{r}_l) - \boldsymbol{r}_j - \boldsymbol{r}_i$, $\boldsymbol{r}_i$ is the position of the localized orbitals in the unit cell. Band structure of monolayer GeS with and without symmetrization of the Hamiltonian is shown in Supplementary Figure 6.

**Data availability**

The data that support the findings of this study are available from the corresponding author upon reasonable request.

**Acknowledgements**

This work was supported by the National Science Foundation under Award #DMR-1753054 and #DMR-2103842, and by an Office of Naval Research MURI through grant #N00014-17-1-2661. We also acknowledge the advanced computing resources provided by Texas A&M High Performance Research Computing.

**Author contributions**

X.Q. conceived the project. X.Q. and J.L. supervised the project. H.W. and X.Q. developed the formula. H.W. developed first-principles code for Wilson loop method and carried out the calculations. H.W. and X.Q. wrote the manuscript with the help of others. All authors conducted theoretical analysis and analyzed the results.

**Competing interests**

The authors declare no competing interests.

**Additional information**

Supplementary information is available.

# Supplementary Information

# Generalized Wilson loop method for nonlinear light-matter interaction

Hua Wang,[1,2,3] Xiuyu Tang,[1] Haowei Xu,[3] Ju Li,[3,4]* and Xiaofeng Qian[1,5]*

[1]Department of Materials Science and Engineering, Texas A&M University, College Station, TX 77843, USA

[2]Future Science Research Institute, ZJU-Hangzhou Global Scientific and Technological Innovation Center, Zhejiang University, Hangzhou, Zhejiang 310058, China

[3]Department of Nuclear Science and Engineering, Massachusetts Institute of Technology, Cambridge, Massachusetts 02139, USA

[4]Department of Materials Science and Engineering, Massachusetts Institute of Technology, Cambridge, Massachusetts 02139, USA

[5]Department of Electrical and Computer Engineering, Texas A&M University, College Station, TX 77843, USA

*Correspondence to: liju@mit.edu, feng@tamu.edu

**Supplementary Note 1. Equivalence of our Wilson loop approach and Young and Rappe's formula**

Shift vector is defined as

$$R_{mn}^{a,b}(\boldsymbol{k}) = -\partial_{k_a}\phi_{mn}^{b}(\boldsymbol{k}) + \mathcal{A}_m^a(\boldsymbol{k}) - \mathcal{A}_n^a(\boldsymbol{k}). \tag{1}$$

The intraband Berry connection reads $\boldsymbol{\mathcal{A}}_m(\boldsymbol{k}) = i\langle m,\boldsymbol{k}|\partial_{\boldsymbol{k}}|m,\boldsymbol{k}\rangle = \lim_{\boldsymbol{q}\to 0}\partial_{\boldsymbol{q}}\mathrm{Im}\ln\langle m,\boldsymbol{k}+\boldsymbol{q}|m,\boldsymbol{k}\rangle$. In contrast, the interband Berry connection between states $|m\rangle$ and $|n\rangle$ is given by $\boldsymbol{r}_{mn}(\boldsymbol{k}) = i\langle n|\partial_{\boldsymbol{k}}|m\rangle = |\boldsymbol{r}_{mn}(\boldsymbol{k})|e^{i\phi_{mn}}$, where $\phi_{mn}$ is the phase of interband Berry connection: $\phi_{mn}^{b}(\boldsymbol{k}) = \mathrm{Im}\ln\left(r_{mn}^{b}(\boldsymbol{k})\right)$. For small $\boldsymbol{q}$, $\langle m,\boldsymbol{k}|m,\boldsymbol{k}+\boldsymbol{q}\rangle = e^{-i\boldsymbol{q}\cdot\boldsymbol{\mathcal{A}}_m(\mathbf{k})+O(q^2)}$, and $\langle m,\boldsymbol{k}+\boldsymbol{q}|m,\boldsymbol{k}\rangle = e^{i\boldsymbol{q}\cdot\boldsymbol{\mathcal{A}}_m(\mathbf{k})+O(q^2)}$. Thus, shift vector can be reformulated as

$$R_{mn}^{a,b}(\boldsymbol{k}) = -\partial_{k_a}\mathrm{Im}\ln\left(r_{mn}^{b}(\boldsymbol{k})\right) + \mathcal{A}_m^a(\boldsymbol{k}) - \mathcal{A}_n^a(\boldsymbol{k}). \tag{2}$$

Thus,

$$R_{mn}^{a,b}(\boldsymbol{k}) = -\partial_{k_a}\mathrm{Im}\ \ln\left(r_{mn}^{b}(\boldsymbol{k})\right) - \lim_{q_a\to 0}\partial_{q_a}\mathrm{Im}\ \ln\langle m,\boldsymbol{k}|m,\boldsymbol{k}+\boldsymbol{q}_a\rangle\langle n,\boldsymbol{k}+\boldsymbol{q}_a|n,\boldsymbol{k}\rangle. \tag{3}$$

Since $\partial_{k_a}\mathrm{Im}\ \ln\left(r_{mn}^{b}(\boldsymbol{k})\right) = \lim_{q_a\to 0}\partial_{q_a}\mathrm{Im}\ \ln\langle m,\boldsymbol{k}+\boldsymbol{q}_a|r^b|n,\boldsymbol{k}+\boldsymbol{q}_a\rangle$,

$$R_{mn}^{a,b}(\mathbf{k}) = -\lim_{q_a\to 0}\partial_{q_a}\mathrm{Im}\ \ln\langle m,\boldsymbol{k}|m,\boldsymbol{k}+\boldsymbol{q}_a\rangle\langle m,\boldsymbol{k}+\boldsymbol{q}_a|r^b|n,\boldsymbol{k}+\boldsymbol{q}_a\rangle\langle n,\boldsymbol{k}+\boldsymbol{q}_a|n,\boldsymbol{k}\rangle. \tag{4}$$

Since $\langle n,\boldsymbol{k}|r^b|m,\boldsymbol{k}\rangle$ does not depend on $\boldsymbol{q}$, we can rewrite $R_{mn}^{a,b}(\mathbf{k})$ as Wilson loop as follows,

$$R_{mn}^{a,b}(\boldsymbol{k}) = -\lim_{q_a\to 0}\partial_{q_a}\mathrm{Im}\ \ln\langle m,\boldsymbol{k}|m,\boldsymbol{k}+\boldsymbol{q}_a\rangle\langle m,\boldsymbol{k}+\boldsymbol{q}_a|r^b|n,\boldsymbol{k}+\boldsymbol{q}_a\rangle\langle n,\boldsymbol{k}+\boldsymbol{q}_a|n,\boldsymbol{k}\rangle\langle n,\boldsymbol{k}|r^b|m,\boldsymbol{k}\rangle. \tag{5}$$

Next let's revisit Eq. (3). The first term can be rewritten explicitly as follows

$$\partial_{k_a}\mathrm{Im}\ln\left(r_{mn}^{b}(\boldsymbol{k})\right) = \lim_{q_a\to 0}\partial_{q_a}\mathrm{Im}\ln\langle m,\boldsymbol{k}+\boldsymbol{q}_a|r^b|n,\boldsymbol{k}+\boldsymbol{q}_a\rangle. \tag{6}$$

Taking the first order Taylor expansion of Bloch states

$$|n,\boldsymbol{k}+\boldsymbol{q}_a\rangle = |n,\boldsymbol{k}\rangle + \left(\partial_{k_a}|n,\boldsymbol{k}\rangle\right)\cdot q_a, \tag{7}$$

$$\langle m,\boldsymbol{k}+\boldsymbol{q}_a| = \langle m,\boldsymbol{k}| + \left(\langle m,\boldsymbol{k}|\partial_{k_a}\right)\cdot q_a, \tag{8}$$

we then have

$$\lim_{q_a\to 0}\partial_{q_a}\ln\langle m,\boldsymbol{k}+\boldsymbol{q}_a|r^b|n,\boldsymbol{k}+\boldsymbol{q}_a\rangle = \frac{\ln\ \langle m,\boldsymbol{k}|r^b\left(\partial_{k_a}|n,\boldsymbol{k}\rangle\right) + \ln\ \left(\langle m,\boldsymbol{k}|\partial_{k_a}\right)r^b|n,\boldsymbol{k}\rangle}{\langle m,\boldsymbol{k}|r^b|n,\boldsymbol{k}\rangle}. \tag{9}$$

Following the strategy of King-Smith and Vanderbilt [1], we obtain the discretized expression

$$\begin{aligned}\langle m,\boldsymbol{k}|\ r^b\left(\partial_{k_a}|n,\boldsymbol{k}\rangle\right) &= \langle m,\boldsymbol{k}|r^b|n,\boldsymbol{k}\rangle\frac{\partial}{\partial k_a'}\Big|_{k_a'=k_a}\ln\langle m,\boldsymbol{k}|r^b|n,\boldsymbol{k}'\rangle\\ &= \langle m,\boldsymbol{k}|r^b|n,\boldsymbol{k}\rangle\lim_{q_a\to 0}\frac{1}{q_a}\left(\ln\langle m,\boldsymbol{k}|r^b|n,\boldsymbol{k}+\boldsymbol{q}_a\rangle - \ln\langle m,\boldsymbol{k}|r^b|n,\boldsymbol{k}\rangle\right)\\ &= \langle m,\boldsymbol{k}|r^b|n,\boldsymbol{k}\rangle\lim_{q_a\to 0}\frac{1}{q_a}\left(\ln\frac{\langle m,\boldsymbol{k}|r^b|n,\boldsymbol{k}+\boldsymbol{q}_a\rangle}{\langle m,\boldsymbol{k}|r^b|n,\boldsymbol{k}\rangle}\right)\end{aligned} \tag{10}$$

Hence, we arrive at the shift vector formula

$$\begin{aligned}R_{mn}^{a,b}(\boldsymbol{k}) &= -\partial_{k_a}\mathrm{Im}\ln\left(r_{mn}^{b}(\boldsymbol{k})\right) + \mathcal{A}_m^a(\boldsymbol{k}) - \mathcal{A}_n^a(\boldsymbol{k})\\ &= \lim_{q_a\to 0}\frac{1}{q_a}\mathrm{Im}\left(\ln\frac{\langle m,\boldsymbol{k}|r^b|n,\boldsymbol{k}\rangle\langle n,\boldsymbol{k}|n,\boldsymbol{k}+\boldsymbol{q}_a\rangle}{\langle m,\boldsymbol{k}|r^b|n,\boldsymbol{k}+\boldsymbol{q}_a\rangle} + \ln\frac{\langle m,\boldsymbol{k}|r^b|n,\boldsymbol{k}\rangle}{\langle m,\boldsymbol{k}|m,\boldsymbol{k}+\boldsymbol{q}_a\rangle\langle m,\boldsymbol{k}+\boldsymbol{q}_a|r^b|n,\boldsymbol{k}\rangle}\right),\end{aligned} \tag{11}$$

which recovers Young and Rappe's formula[2]. We have also verified that the two shift current expressions are equavilent numerically.

**Supplementary Note 2. Derivation of covariant generalized derivative**

With

$$\langle n,\boldsymbol{k}|[\boldsymbol{r}_i,\boldsymbol{O}]|m,\boldsymbol{k}'\rangle = i\delta(\boldsymbol{k}-\boldsymbol{k}')(\boldsymbol{O}_{mn})_{;\boldsymbol{k}}, \tag{12}$$

$$(\boldsymbol{O}_{nm})_{;\boldsymbol{k}} \equiv \frac{\partial \boldsymbol{O}_{nm}}{\partial \boldsymbol{k}} - i\boldsymbol{O}_{nm}(\boldsymbol{\mathcal{A}}_n - \boldsymbol{\mathcal{A}}_m), \tag{13}$$

we expand the commutator as follows

$$\begin{aligned}\langle n,\boldsymbol{k}|[\boldsymbol{r}_i,\boldsymbol{O}]|m,\boldsymbol{k}'\rangle &= \langle n,\boldsymbol{k}|\boldsymbol{r}_i\boldsymbol{O}-\boldsymbol{O}\boldsymbol{r}_i|m,\boldsymbol{k}'\rangle \\ &= \sum_{l,\boldsymbol{k}''}\langle n,\boldsymbol{k}|\boldsymbol{r}_i|l,\boldsymbol{k}''\rangle\langle l,\boldsymbol{k}''|\boldsymbol{O}|m,\boldsymbol{k}'\rangle - \sum_{l,\boldsymbol{k}''}\langle n,\boldsymbol{k}|\boldsymbol{O}|l,\boldsymbol{k}''\rangle\langle l,\boldsymbol{k}''|\boldsymbol{r}_i|m,\boldsymbol{k}'\rangle.\end{aligned} \tag{14}$$

We then rewrite the first term

$$\begin{aligned}\sum_{l,\boldsymbol{k}''}\langle n,\boldsymbol{k}|\boldsymbol{r}_i|l,\boldsymbol{k}''\rangle\langle l,\boldsymbol{k}''|\boldsymbol{O}|m,\boldsymbol{k}'\rangle &= \sum_{l,\boldsymbol{k}''}\delta_{nl}[\delta(\boldsymbol{k}-\boldsymbol{k}'')\boldsymbol{\mathcal{A}}_n + i\nabla_{\boldsymbol{k}}\delta(\boldsymbol{k}-\boldsymbol{k}'')]\delta(\boldsymbol{k}'-\boldsymbol{k}'')\boldsymbol{O}_{lm}(\boldsymbol{k}') \\ &= \sum_{\boldsymbol{k}''}[\delta(\boldsymbol{k}-\boldsymbol{k}'')\boldsymbol{\mathcal{A}}_n + i\nabla_{\boldsymbol{k}}\delta(\boldsymbol{k}-\boldsymbol{k}'')]\delta(\boldsymbol{k}'-\boldsymbol{k}'')\boldsymbol{O}_{nm}(\boldsymbol{k}') \\ &= \sum_{\boldsymbol{k}''}\delta(\boldsymbol{k}-\boldsymbol{k}'')\delta(\boldsymbol{k}'-\boldsymbol{k}'')\boldsymbol{\mathcal{A}}_n\boldsymbol{O}_{nm}(\boldsymbol{k}') + i\sum_{\boldsymbol{k}''}[\nabla_{\boldsymbol{k}}\delta(\boldsymbol{k}-\boldsymbol{k}'')]\delta(\boldsymbol{k}'-\boldsymbol{k}'')\boldsymbol{O}_{nm}(\boldsymbol{k}') \\ &= \sum_{\boldsymbol{k}''}\delta(\boldsymbol{k}-\boldsymbol{k}'')\delta(\boldsymbol{k}'-\boldsymbol{k}'')\boldsymbol{\mathcal{A}}_n\boldsymbol{O}_{nm}(\boldsymbol{k}') + i\sum_{\boldsymbol{k}''}[\nabla_{\boldsymbol{k}}\delta(\boldsymbol{k}-\boldsymbol{k}'')]\delta(\boldsymbol{k}'-\boldsymbol{k}'')\boldsymbol{O}_{nm}(\boldsymbol{k}') \\ &= \sum_{\boldsymbol{k}''}\delta(\boldsymbol{k}-\boldsymbol{k}'')\delta(\boldsymbol{k}'-\boldsymbol{k}'')\boldsymbol{\mathcal{A}}_n\boldsymbol{O}_{nm}(\boldsymbol{k}') + i[\nabla_{\boldsymbol{k}}\delta(\boldsymbol{k}-\boldsymbol{k}')]\boldsymbol{O}_{nm}(\boldsymbol{k}') \\ &= \delta(\boldsymbol{k}-\boldsymbol{k}')\boldsymbol{\mathcal{A}}_n\boldsymbol{O}_{nm}(\boldsymbol{k}') + i\nabla_{\boldsymbol{k}}[\delta(\boldsymbol{k}-\boldsymbol{k}')]\boldsymbol{O}_{nm}(\boldsymbol{k}') \\ &= \delta(\boldsymbol{k}-\boldsymbol{k}')\boldsymbol{\mathcal{A}}_n\boldsymbol{O}_{nm}(\boldsymbol{k}') + i\nabla_{\boldsymbol{k}}[\delta(\boldsymbol{k}-\boldsymbol{k}')\boldsymbol{O}_{nm}(\boldsymbol{k}')].\end{aligned} \tag{15}$$

Using the same strategy, we rewrite the second term

$$\begin{aligned}\sum_{l,\boldsymbol{k}''}\langle n,\boldsymbol{k}|\boldsymbol{O}|l,\boldsymbol{k}''\rangle\langle l,\boldsymbol{k}''|\boldsymbol{r}_i|m,\boldsymbol{k}'\rangle &= \sum_{l,\boldsymbol{k}''}\delta(\boldsymbol{k}-\boldsymbol{k}'')\boldsymbol{O}_{nl}(\boldsymbol{k}'')\delta_{lm}[\delta(\boldsymbol{k}'-\boldsymbol{k}'')\boldsymbol{\mathcal{A}}_m + i\nabla_{\boldsymbol{k}''}\delta(\boldsymbol{k}'-\boldsymbol{k}'')] \\ &= \sum_{\boldsymbol{k}''}\delta(\boldsymbol{k}-\boldsymbol{k}'')\boldsymbol{O}_{nm}(\boldsymbol{k}'')[\delta(\boldsymbol{k}'-\boldsymbol{k}'')\boldsymbol{\mathcal{A}}_m + i\nabla_{\boldsymbol{k}''}\delta(\boldsymbol{k}'-\boldsymbol{k}'')] \\ &= \delta(\boldsymbol{k}-\boldsymbol{k}')\boldsymbol{\mathcal{A}}_m\boldsymbol{O}_{nm}(\boldsymbol{k}') + \sum_{\boldsymbol{k}''}\delta(\boldsymbol{k}-\boldsymbol{k}'')\boldsymbol{O}_{nm}(\boldsymbol{k}'')[i\nabla_{\boldsymbol{k}''}\delta(\boldsymbol{k}'-\boldsymbol{k}'')] \\ &= \delta(\boldsymbol{k}-\boldsymbol{k}')\boldsymbol{\mathcal{A}}_m\boldsymbol{O}_{nm}(\boldsymbol{k}') + \boldsymbol{O}_{nm}(\boldsymbol{k})[i\nabla_{\boldsymbol{k}}\delta(\boldsymbol{k}'-\boldsymbol{k})].\end{aligned} \tag{16}$$

We also have

$$i\nabla_{\boldsymbol{k}}[\delta(\boldsymbol{k}'-\boldsymbol{k})\boldsymbol{O}_{nm}(\boldsymbol{k})] = i\boldsymbol{O}_{nm}(\boldsymbol{k})\nabla_{\boldsymbol{k}}\delta(\boldsymbol{k}'-\boldsymbol{k}) + i\delta(\boldsymbol{k}'-\boldsymbol{k})\nabla_{\boldsymbol{k}}\boldsymbol{O}_{nm}(\boldsymbol{k}), \tag{17}$$

$$i\boldsymbol{O}_{nm}(\boldsymbol{k})\nabla_{\boldsymbol{k}}\delta(\boldsymbol{k}'-\boldsymbol{k}) = i\nabla_{\boldsymbol{k}}[\delta(\boldsymbol{k}'-\boldsymbol{k})\boldsymbol{O}_{nm}(\boldsymbol{k})] - i\delta(\boldsymbol{k}'-\boldsymbol{k})\nabla_{\boldsymbol{k}}\boldsymbol{O}_{nm}(\boldsymbol{k}). \tag{18}$$

Using the relation

$$\delta(\boldsymbol{k}-\boldsymbol{k}')\boldsymbol{O}_{nm}(\boldsymbol{k}') = \delta(\boldsymbol{k}'-\boldsymbol{k})\boldsymbol{O}_{nm}(\boldsymbol{k}), \tag{19}$$

we finally obtain

$$\langle n,\boldsymbol{k}|[\boldsymbol{r}_i,\boldsymbol{O}]|m,\boldsymbol{k}'\rangle = i\delta(\boldsymbol{k}-\boldsymbol{k}')\nabla_{\boldsymbol{k}}\boldsymbol{O}_{nm}(\boldsymbol{k}) - i\delta(\boldsymbol{k}-\boldsymbol{k}')\boldsymbol{O}_{nm}(\boldsymbol{\mathcal{A}}_n - \boldsymbol{\mathcal{A}}_m). \tag{20}$$

**III. Supplementary Figures**

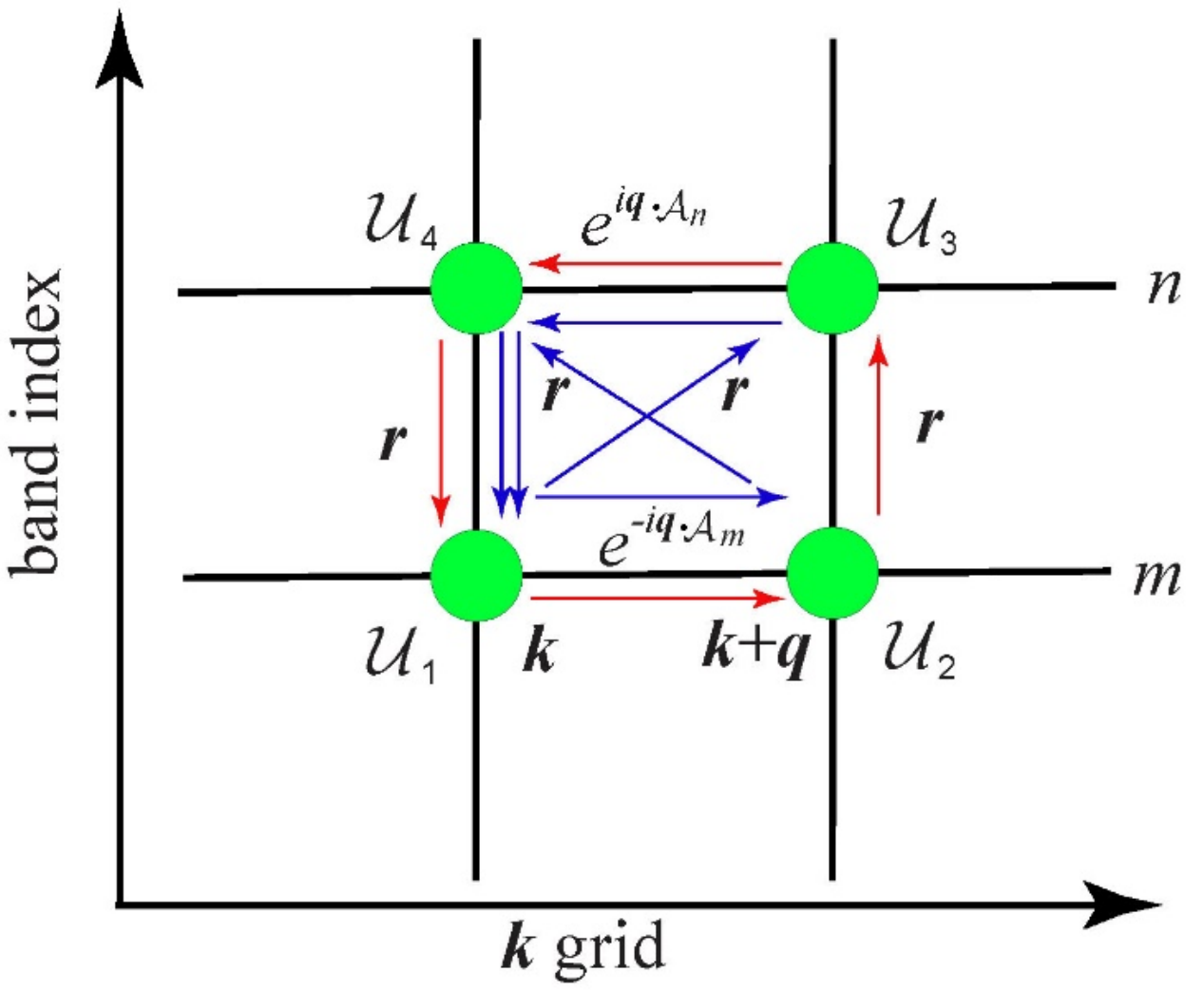

**Supplementary Figure 1.** Two Wilson loop representations of shift vector. The first loop denoted by red arrows was used in the main text, which reads $W_{mn}(\boldsymbol{k}) = \langle\mathcal{U}_1|\mathcal{U}_2\rangle\langle\mathcal{U}_2|\boldsymbol{r}|\mathcal{U}_3\rangle\langle\mathcal{U}_3|\mathcal{U}_4\rangle\langle\mathcal{U}_4|\boldsymbol{r}|\mathcal{U}_1\rangle$. The loop indicated by blue arrows reads $W_{mn}(\boldsymbol{k}) = \langle\mathcal{U}_1|\boldsymbol{r}|\mathcal{U}_3\rangle\langle\mathcal{U}_3|\mathcal{U}_4\rangle\langle\mathcal{U}_4|\boldsymbol{r}|\mathcal{U}_1\rangle\langle\mathcal{U}_1|\mathcal{U}_2\rangle\langle\mathcal{U}_2|\boldsymbol{r}|\mathcal{U}_4\rangle\langle\mathcal{U}_4|\boldsymbol{r}|\mathcal{U}_1\rangle$.

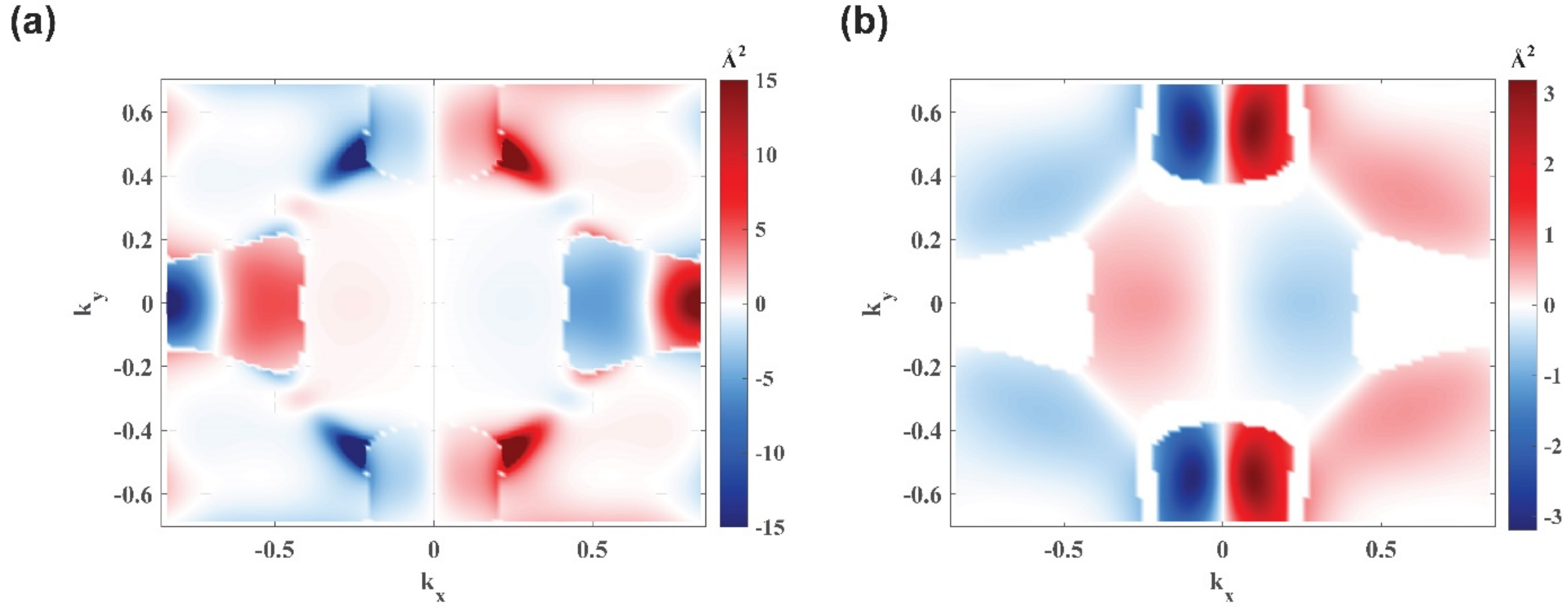

**Supplementary Figure 2.** (a) Intraband Berry curvature of the lowest conduction band in the first Brillouin zone using the Wilson loop approach for monolayer GeS. (b) Calculated interband Berry curvature across the gap in the first Brillouin zone using the Wilson loop approach for monolayer GeS.

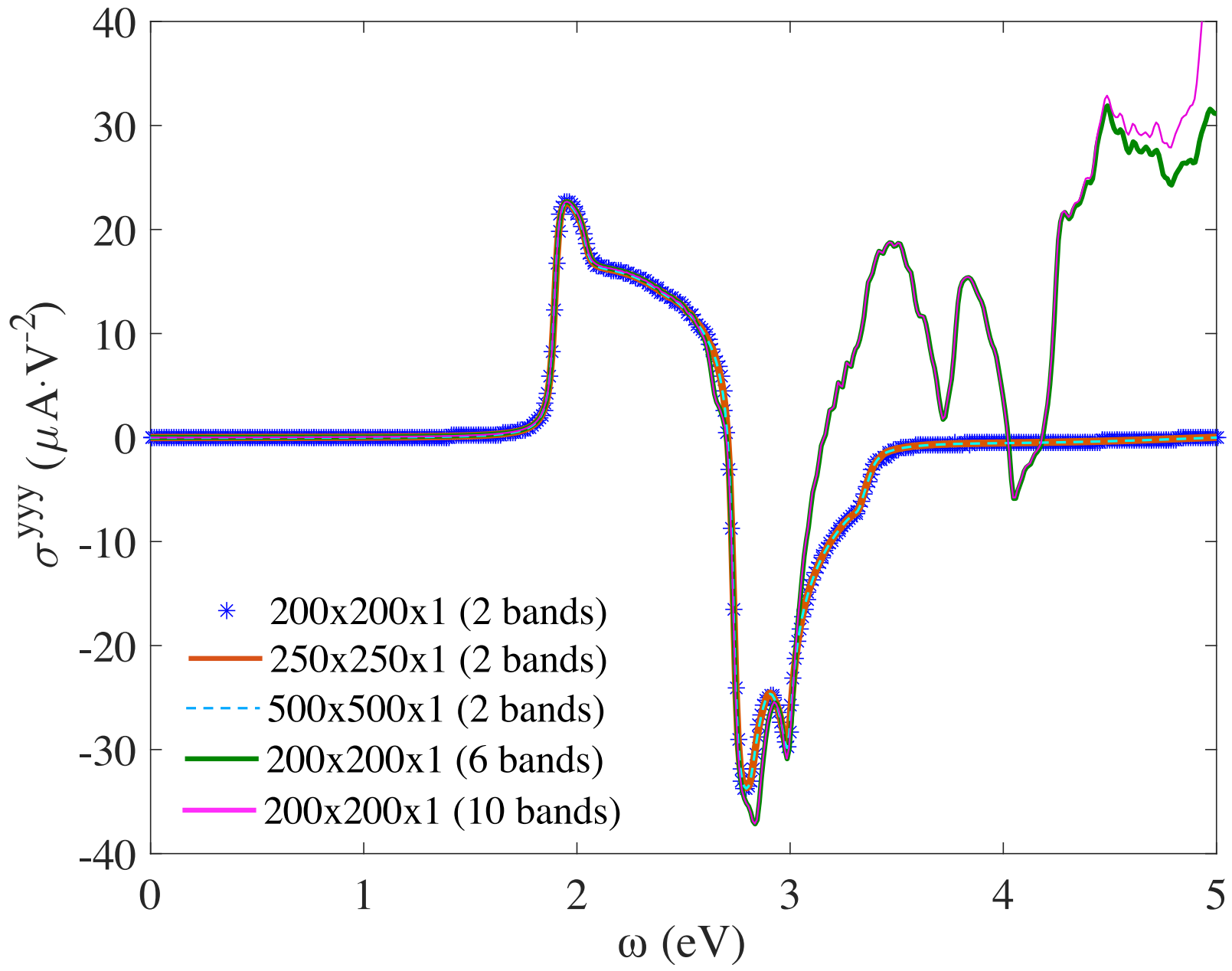

**Supplementary Figure 3.** Shift current conductivity with different number of bands and ***k***-point sampling. The results are converged with ***k***-points of 250×250×1 and 500×500×1 and converged with two bands for frequencies less than ~3 eV.

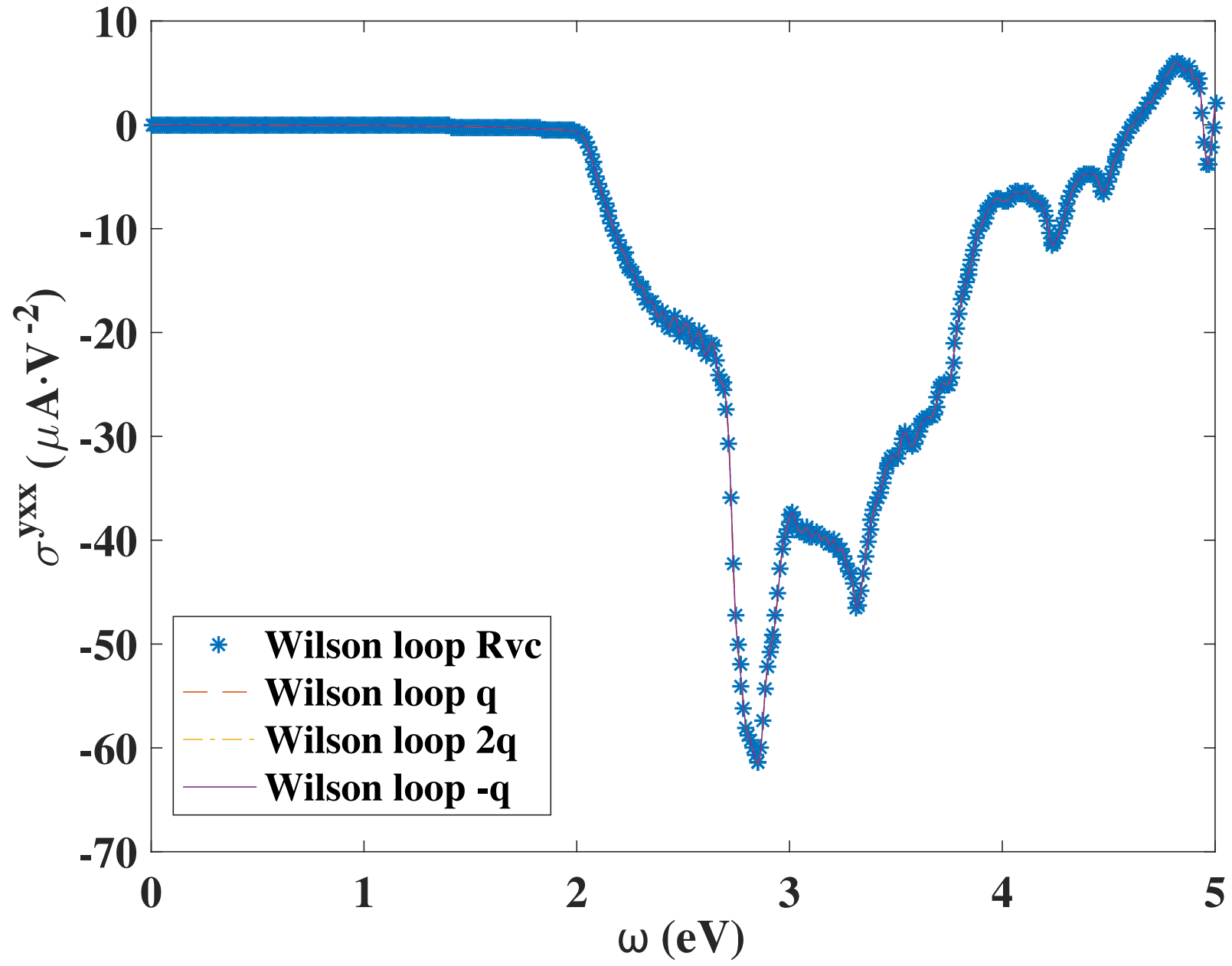

**Supplementary Figure 4.** Calculated shift current conductivity $\sigma^{yxx}(\omega)$ by the Wilson loop method with two bands across the gap. *R*vc denotes the original shift vector formula. q, 2q, and -q represent the numerical calculation of the imaginary part of Wilson loop with different q values.

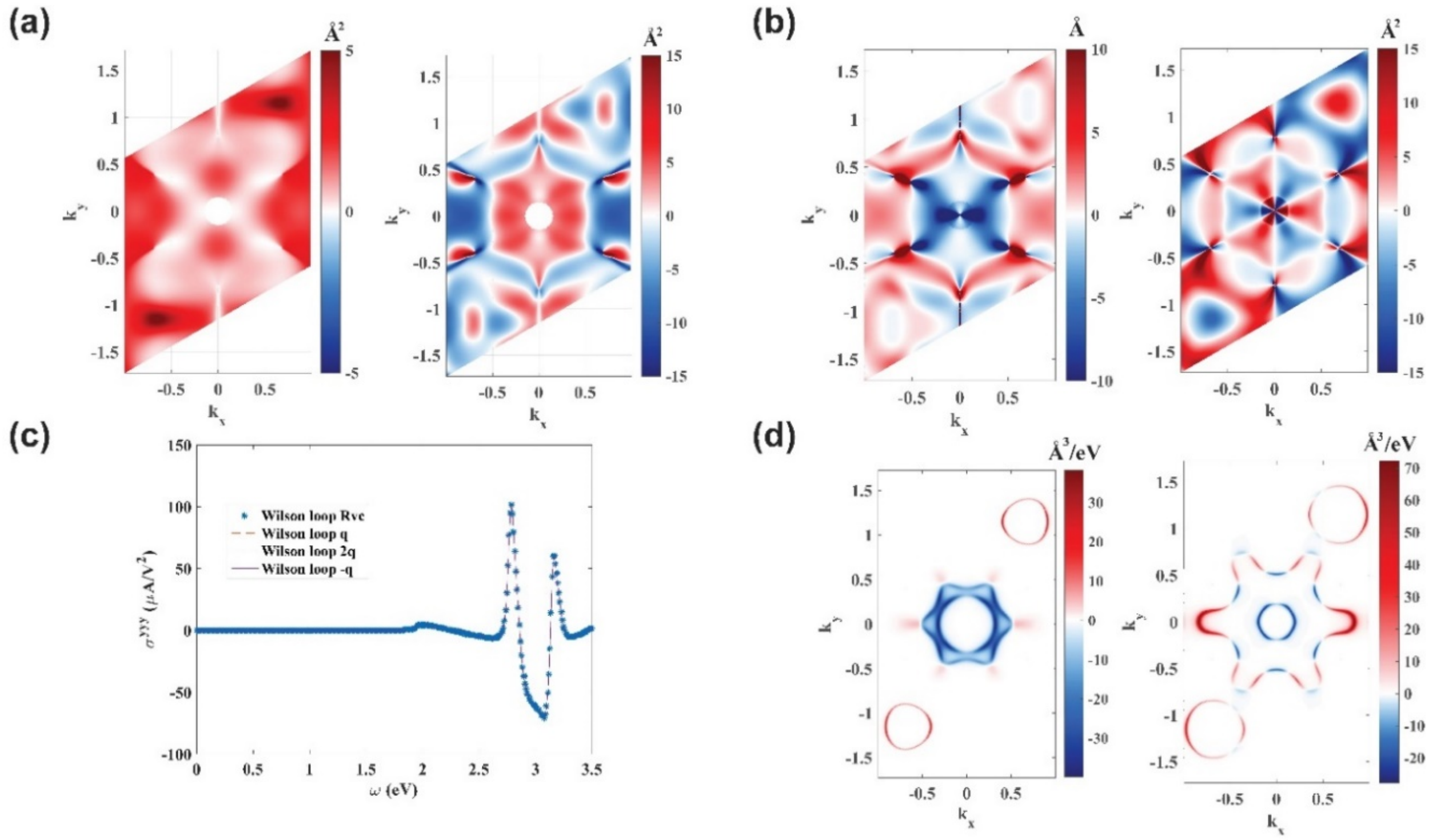

**Supplementary Figure 5.** Nonlinear optical responses of monolayer $WS_2$ with the generalized Wilson loop approach. (a) Calculated $\boldsymbol{k}$-resolved real and imaginary part of the Wilson loop Re $W_{mn}\left(\boldsymbol{k}, \boldsymbol{q_y} \to 0, r^y, r^y\right)$ and $\lim_{q_a \to 0} \frac{1}{q_a} \mathrm{Im}\, W_{mn}\left(\boldsymbol{k}, \boldsymbol{q_y} \to 0, r^y, r^y\right)$ with two bands across the gap. (b)$\boldsymbol{k}$-resolved shift vector and Berry curvature. **c** Shift current conductivity tensor $\sigma^{yyy}$ with six bands. (c, d) $k$-resolved shift current strength at $\omega = 2.8$ and $\omega = 3.0$ eV using the generalized Wilson loop approach, respectively.

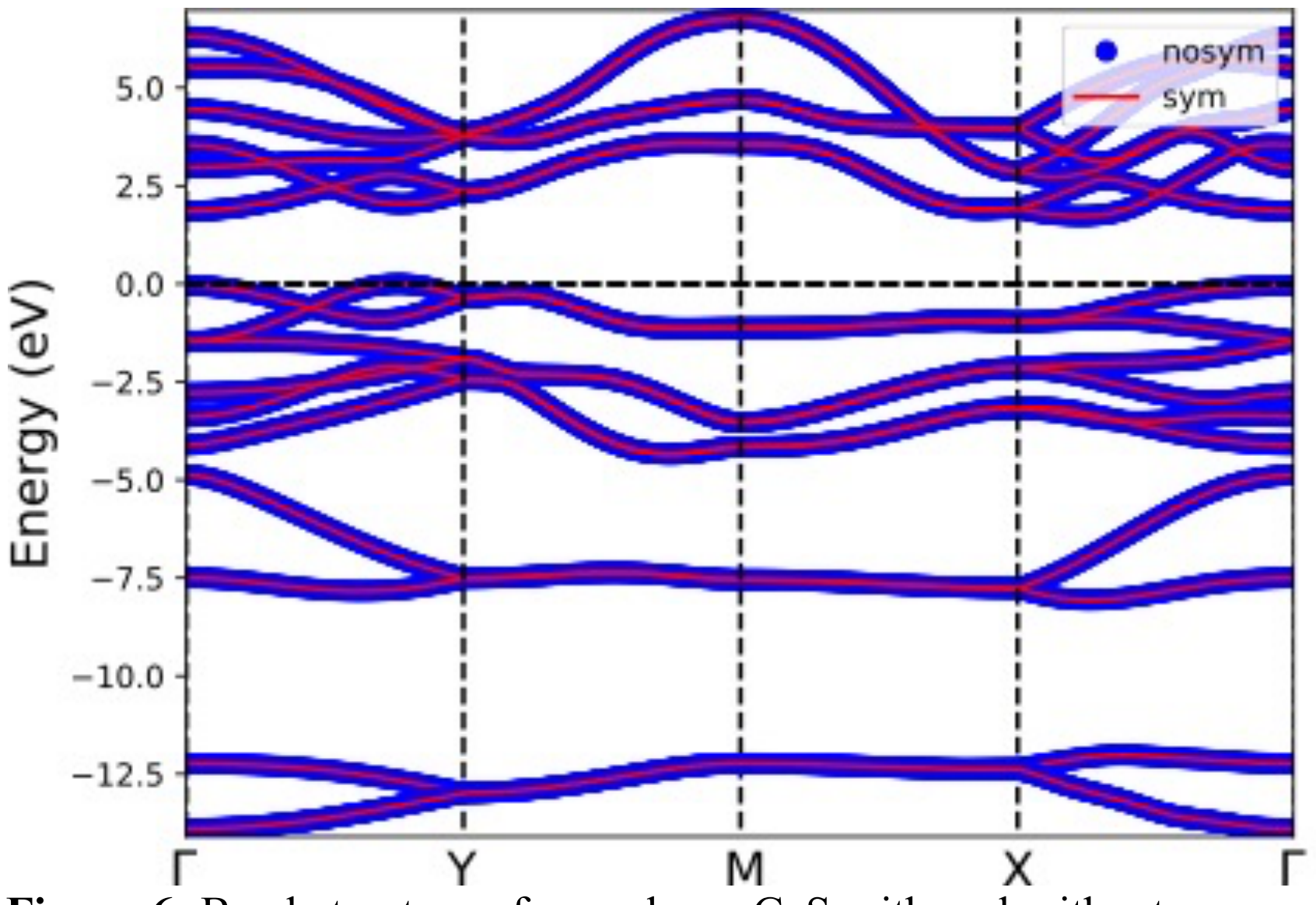

**Supplementary Figure 6.** Band structure of monolayer GeS with and without symmetrization of tight-binding Hamiltonian.

## Supplementary References